\documentclass[amssymb, reprint, nobibnotes, showpacs, superscriptaddress, aps, pra]{revtex4-1}
\usepackage{graphicx,amsmath}

\begin{document}

\title{Dynamical Properties of Three-Photon Blockade in a Two-Atom Cavity QED System}

\author{DaQiang Bao}
\affiliation{MOE Key Laboratory of Advanced Micro-Structured Materials, School of Physics Science and Engineering, Tongji University, Shanghai 200092, China}
\author{Chengjie Zhu}
\affiliation{MOE Key Laboratory of Advanced Micro-Structured Materials, School of Physics Science and Engineering, Tongji University, Shanghai 200092, China}
\email[]{cjzhu@tongji.edu.cn}
\author{Jingping Xu}
\affiliation{MOE Key Laboratory of Advanced Micro-Structured Materials, School of Physics Science and Engineering, Tongji University, Shanghai 200092, China}
\author{Yaping Yang}
\affiliation{MOE Key Laboratory of Advanced Micro-Structured Materials, School of Physics Science and Engineering, Tongji University, Shanghai 200092, China}
\email[]{yang\_yaping@tongji.edu.cn}

\date{\today}




\begin{abstract}
We theoretically investigate the three-photon blockade phenomenon in a two atoms cavity QED system, where two atoms are driven by a coherent field. In the case of in-phase radiations, we show that the three-photon blockade with bunched two photons can be realized in a small regime of the driving field Rabi frequency. However, in the case of out-of-phase radiations, the one-photon excitations are prohibited due to the destructive interference, and three-photon blockade can be realized in a wide regime of driving field Rabi frequency. In addition, the three-photon blockade phenomenon can be significantly improved. The results presented here show that this two-atom scheme is a good candidate to achieve anti-bunched photon pairs.
\end{abstract}


\maketitle

\section{Introduction.}
Nonclassical light is one of the research hotspots in quantum optics and its  statistical properties can be described by the sub-Poissonian distribution and anti-bunching behavior~\cite{mandel}. Usually, the sub-Poissonian distribution is described as a distribution that the variance of the photon number is less than the Poissonian distribution, while the anti-bunching behavior is defined as a phenomenon that photon tend to be apart from one another~\cite{scully}. The nonclassical light can be generated in many systems via the interaction between light and matter. In particular, the cavity quantum electrodynamics (QED) system is widely used to generate nonclassical light via the photon blockade effect.

In a strongly coupled atom-cavity system, it is impossible to realize that  both single-photon absorption and two-photon absorption are resonant at same time due to the dressed states anharmonicity~\cite{agarwal}. This inhibition of resonant absorption of the second photon is known as the conventional two-photon blockade, which was firstly observed by Birnbaum et al.~\cite{KM}. They show that a coherent probe field was converted into a non-classical field, and the statistics of output light satisfied the sub-Poissonian distribution and anti-bunching character. Using the two-photon blockade effect, one can achieve the output of single photon in cavity QED systems. Since then, a lot of interest has been payed in the research of photon blockade due to its potential applications in quantum communication and quantum information processing~\cite{p1,p2,p3,p4,p5,p6}. 

Using the quantum interference between excitation pathways, the unconventional photon blockade was proposed to improve the two-photon blockade effect~\cite{u2}. To date, many theoretical schemes have been proposed~\cite{up1,up2,up3,up4,up5}, and the observations of unconventional photon blockade were recently demonstrated by Snijders et al.~\cite{un1} and Vaneph et al.~\cite{un2}. Although the two-photon blockade has been widely studied, the research work on the three-photon blockade (i.e., the blockade of absorption of the third photon) is scarce. 
This is because that a strong pump field is required to generate the two-photon excitations and the power broadening compensate the dressed state anharmonicity. Therefore, the third photon can not be blockaded efficiently and it is a challenge to realize three-photon blockade in a typical single atom-cavity QED system~\cite{hamsen}.

In this paper, we consider that two two-level atoms are trapped in a single mode cavity, where the one-photon excitations are forbidden due to the destructive interference, and two-photon excitations are dominant under the case of atom driving and out-of-phase radiations~\cite{zhu}. As a result, it is easy to realize three-photon blockade in this two atoms cavity QED system. Here, we investigate the dynamical properties of the three photon blockade phenomenon and the statistical characteristics of the photon distributions. We show that the phase between two atoms is a critical factor to realize the three-photon blockade with anti-bunched photon pairs. When two atoms radiate in-phase, the three photon blockade can only be realized in a small regime of the driving field Rabi frequency. However, in the case of out-of-phase radiations, the three-photon blockade can be realized in a wide regime of the driving field intensity. Moreover, the three photon blockade can be significantly improved in the case of out-phase radiations. 

\begin{figure}[htbp]
\centering
\includegraphics[width=0.5\textwidth]{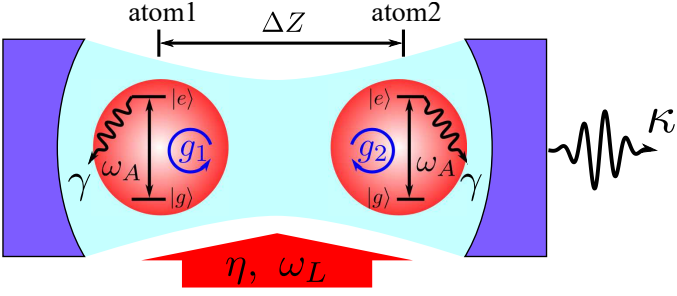}
\caption{Sketch of two atoms cavity QED system. ${g}_i\ (i=1,2)$ is the coupling strength for the $i$-th two-level atom with identical resonant frequency $\omega_A$, and $\Delta z$ is the distance between two atoms. The system is driven by a coherent pump field with Rabi frequency $\eta$. $\gamma$ and $\kappa$ are the decay rate of the atom and cavity, respectively.}~\label{model}
\end{figure} 
First, we consider that two identical two-level atoms are trapped in a single mode cavity with resonant frequency $\omega_{\rm cav}$. As shown in Fig.~1, the ground and excited states of each atom are labeled as $|g\rangle$ and $|e\rangle$, respectively. These two atoms with the same resonant frequency $\omega_A$ and raising (lowering) operator $\sigma_i^{\pm}\ (i=1,2)$ are separated by distance $\Delta z$. A pump field with the angular frequency $\omega_L$ and Rabi frequency $\eta$ drives these two atoms directly in this cavity QED system. Under the electric dipole and rotating wave approximations, the dynamics of the system can be described by a master equation of Lindblad form~\cite{zhu}, 
\begin{eqnarray}\label{eq:ME}
\frac{\partial}{\partial t}\rho &=&-i[H,\rho]+\kappa(2a\rho a^\dag-a^\dag a\rho-\rho a^\dag a)\nonumber\\
& &+\gamma\sum\limits_{j=1,2}(2\sigma^-_j\rho \sigma^+_j-\sigma^+_j\sigma^-_j\rho-\rho\sigma^+_j\sigma^-_j),
\end{eqnarray}
where $\rho$ is the density of the system and the corresponding Hamiltonian $H=\Delta_a\sum\limits_{j=1,2}\sigma^+_j\sigma^-_j+\Delta_{\rm cav}a^\dag a+\eta\sum\limits_j(\sigma_j^++\sigma_j^-)+\sum\limits_{j=1,2}g_j(a\sigma_j^++a^\dag\sigma_j^-)$. 
Here, 
$a$ ($a^\dag$) is the creation (annihilation) operator for the cavity mode. The atomic and cavity detunings are defined as $\Delta_a=\omega_L-\omega_A$ and $\Delta_{\rm cav}=\omega_L-\omega_{\rm cav}$, respectively. The last term in Eq.~(2) describes the energy exchange between atoms and the cavity mode with coupling strengths between the atom and cavity $g_i=g\cos(2\pi z_i/\lambda_{\rm cav})\ (i=1,2)$ where $z_i$ is the position of $i$-th atom and $\lambda_{\rm cav}=\omega_{\rm cav}/c_0$ is the wavelength of cavity mode. For mathematical simplicity, we assume the left atom is located at the anti-node of the wavelength (i.e., $z_1=0$) so that the coupling strengths can be written as $g_1=g$ and $g_2=g\cos\phi_z$ with $\phi_z=2\pi\Delta z/\lambda_{\rm cav}$. $\gamma$ and $\kappa$ are the atomic spontaneous emission rate and the cavity decay rate, respectively. 

\begin{figure}[htbp]
	\centering
	\includegraphics[width=0.5\textwidth]{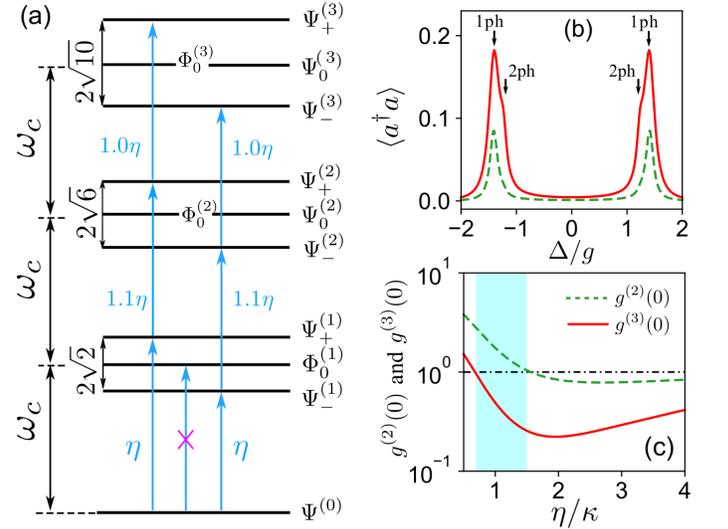}
	\caption{(a) Diagram of eigenstates and the main transition pathways for $\phi_z=0$.
		(b) The mean photon number versus the normalizing detuning $\Delta/g$ with $\eta/\kappa=0.5$ (green line) and $\eta/\kappa=1.0$ (red line). (c) Steady state field correlation functions $g^{(2)}(0)$ (green line) and $g^{(3)}(0)$ (red line) versus the pump field Rabi frequency $\eta$ at two-photon resonance frequency.	  
		}~\label{fig1}
\end{figure}

\section{In phase radiations.}
Considering the case of $\phi_z=0$ (i.e., $g_1=g_2=g$), two atoms have the same radiation phases. Using collective states $|gg,n\rangle$, $|\pm\rangle=(|eg,n-1\rangle\pm|ge,n-1\rangle)/\sqrt{2}$ and $|ee,n-2\rangle$ as new basis to rewrite the Hamiltonian of the system, we can obtain
\begin{equation}
H_I=
\begin{bmatrix}
0          & \sqrt{2n}g    & 0 &              0\\
\sqrt{2n}g &      0        & 0 & \sqrt{2(n-1)}g\\
0          & 0             & 0 &              0\\
0          &\sqrt{2(n-1)}g & 0 &              0
\end{bmatrix}
\end{equation} 
under the weak pump approximation, where $n$ is the photon number. It should be notes that the state $|ee,0\rangle$ can't be existed in one-photon space (i.e., $n=1$). 

Solving Eq.~(3), the corresponding eigenvalues and eigenstates can be obtained easily (see Fig.~2(a)). For example, in one-photon space, we can obtain three eigenvalues $E^{(1)}_{\pm}=\pm\sqrt{2}g$ and $E^{(1)}_0=0$. The corresponding eigenstates are given by $\Psi_{\pm}^{(1)}=(-|gg,1\rangle\mp|+,0\rangle)/\sqrt{2}$ and $\Phi_{0}^{(1)}=|-,0\rangle$. As a result, there exist two peaks in the cavity excitation spectrum for weak pump field shown in Fig.~2(b) (greed dashed curve). Here, we choose $\Delta_a=\Delta_{\rm cav}=\Delta$ and other system parameters are given by $\eta/\kappa=0.5$, $\gamma/\kappa=1.0$, $g_1=g_2=g=15\kappa$.
The eigenvalues in the two-photon space (i.e., $n=2$) are given by $E^{(2)}_\pm=\pm\sqrt{6}g$ and $E^{(2)}_{0\pm}=0$ with the corresponding eigenstates $\Psi_{\pm}^{(2)}=\sqrt3/3|gg,2\rangle\pm\sqrt2/2|+,1\rangle+\sqrt6/6|ee,0\rangle$, $\Phi_{0}^{(2)}=(-|gg,2\rangle+\sqrt{2}|ee,0\rangle)/\sqrt{3}$ and $\Psi_{0}^{(2)}=|-,1\rangle$, respectively. It is noted that two-photon excitations can only be observed if the pump field is strong enough [for example, $\eta/\kappa=1$, see red solid curve in Fig.~2(b)]. In three-photon space ($n=3$), we can also find the eigenvalues $E^{(3)}_\pm\pm\sqrt{10}g$ and $E_{0\pm}^{(3)}=0$ with the corresponding eigenstates $\Psi_{\pm}^{(3)}=\sqrt{30}/10|gg,3\rangle\pm\sqrt2/2|+,2\rangle+\sqrt5/5|ee,1\rangle$, $\Phi^{(3)}_{0}=(-\sqrt{2}|gg,3\rangle+\sqrt{3}|ee,1\rangle)/\sqrt{5}$ and $\Psi^{(3)}_{0}=|-,2\rangle$, respectively. 

Due to the energy anharmonicity of the dressed states shown in Fig.~2(a), it is possible to realize two-photon blockade phenomenon when the pump field frequency is tuned to be resonant with the frequency of one-photon excitations, i.e., $\Delta=\sqrt{2}g$~\cite{hamsen}. In particular, choosing a suitable pump field Rabi frequency and tuning the frequency of the pump field to two-photon resonance (i.e., $\Delta=-\sqrt{6}g/2$), one can achieve three-photon blockade in this cavity QED system~\cite{hamsen,zhu}. In this case, the transition pathway $\Psi^{(0)}\rightarrow\Psi_{\pm}^{(1)}\rightarrow\Psi_{\pm}^{(2)}$ is allowed, but the $\Psi_{\pm}^{(2)}\rightarrow\Psi_{\pm}^{(3)}$ transition is forbidden due to the large frequency difference, resulting in the steady-state second-order field correlation function $g^{(2)}(0)=\langle a^\dag a^\dag a a\rangle_{\rm ss}/\langle a^\dag a\rangle_{\rm ss}^2>1$, but the third-order correlation function $g^{(3)}(0)=\langle a^\dag a^\dag a^\dag a a a\rangle_{\rm ss}/\langle a^\dag a\rangle_{\rm ss}^3<1$. As shown in Fig.~2(c), the three-photon blockade can only be achieved in a narrow frequency regime as indicated by the cyan area.


\begin{figure}[htb]
	\centering
	\includegraphics[width=0.5\textwidth]{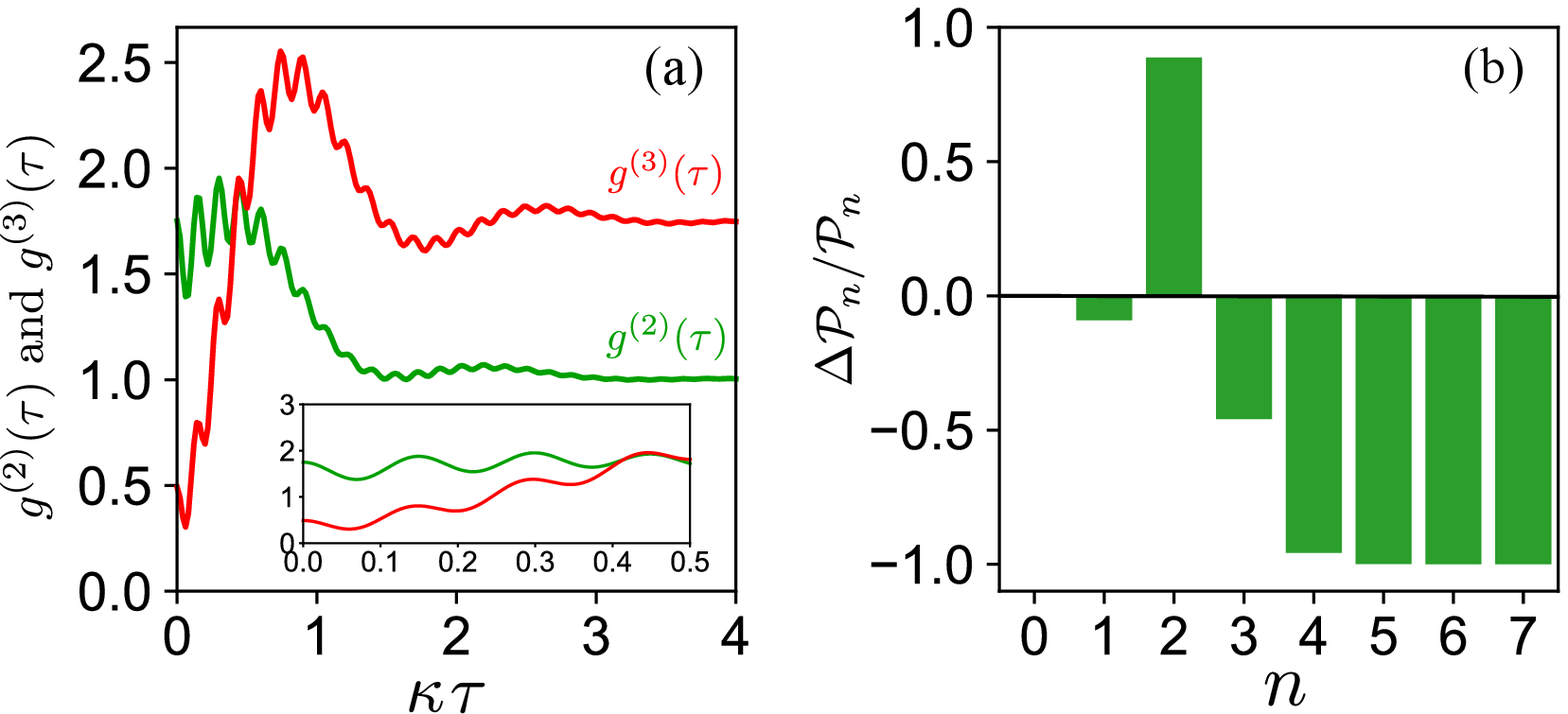}
	\caption{(a) Time dependent field correlation functions $g^{(2)}(\tau)$ (green line) and $g^{(3)}(\tau)$ (red line) versus the normalized delay time $\kappa\tau$ at $\Delta=-\sqrt{6}g/2$ with $\eta/\kappa=1$ and $\phi_z=0$. (b) The deviations of photon number distribution with respect to the Poisson distribution.
		}~\label{fig2}
\end{figure}
In Fig.~3(a), we plot the timed dependent field correlation functions $g^{(2)}(\tau)=\langle a^\dag(t) a^\dag(t+\tau) a(t+\tau) a(t)\rangle/\langle a^\dag(t) a(t)\rangle_{t=\infty}^2$ and $g^{(3)}(\tau)=\langle a^\dag(t) a^\dag(t) a^\dag(t+\tau) a(t+\tau) a(t) a(t)\rangle/\langle a^\dag(t) a(t)\rangle_{t=\infty}^3$ to show the dynamical properties of the three-photon blockade. Here, we choose $\Delta=-\sqrt{6}g/2$, $\eta/\kappa=1$ and other system parameters are the same as those used in Fig.~2(b). Clearly, we have $g^{(2)}(0)\approx1.75>1$ and $g^{(3)}(0)\approx0.5<1$ which implies bunched two photons but anti-bunched photon pairs (i.e., three photon blockade). As the time delay increasing, the second and third-order field correlation functions show many interesting dynamical features.

In general, the second correlation function $g^{2}(\tau)$ is a two-photon measurement. The second manifold eigenstate $\Psi^{(2)}_-$ is dominant if we choose $\Delta=-\sqrt{6}g/2$. After the first photon is measured, the state $\Psi^{(2)}_-=\sqrt3/3|gg,2\rangle-\sqrt2/2|+,1\rangle+\sqrt6/6|ee,0\rangle$ is projected into the state $\Phi_{\rm 1D}=\sqrt6/3|gg,1\rangle-\sqrt2/2|+,0\rangle$ and is not the eigenstate of this system. Apparently, the transition pathway $|gg,1\rangle\overset{g_+}\leftrightarrow|+,0\rangle$ results in the oscillation with the resonant Rabi frequency $2\sqrt{2}g$ (corresponding to time period $T=2\pi/(2\sqrt{2}g)\approx0.14/\kappa$) as shown in Fig.~3(a). It is noted that $g^{(2)}(\tau)$ drops at the beginning of time evolution because the probability of absorbing one photon from the state $|gg,1\rangle$ is larger than that of emitting one photon from the state $|+,0\rangle$.

It is noted that there also exist another coupling pathway without energy exchange by cavity photons, i.e., $|gg,0\rangle\overset{\eta}\leftrightarrow|+,0\rangle$, which results in the oscillation with a lower frequency $\sqrt{4\eta^2+\delta^2}\approx3.5\kappa$. Here, the detuning is defined by $\delta=(\sqrt{2}-\sqrt{6}/2)g$ and the corresponding time period of this slow oscillation is $T=2\pi/\sqrt{4\eta^2+\delta^2}\approx1.8/\kappa$. Therefore, the oscillation in the second order field correlation function originates from two different physical mechanisms. The faster oscillation is the consequence of coherent exchange of a single photon between the atom and the cavity. While, the slower oscillation is due to the exchange of energy between the atom and the external pump field~\cite{koch,koch2}.

To examine the dynamical properties of the third-order field correlation function $g^{(3)}(\tau)$, eigenstates in $n\geq2$ photon space have to be consider. Taking $\Psi_-^{(3)}$ as an example, after two photons have been detected, it is projected into the state $\Psi_{\rm 2D}=3/\sqrt{5}|gg,1\rangle-|+,0\rangle$. Apparently, 
the third-order correlation function also oscillate with two different resonant frequency ($2\sqrt{2}g$ and $\sqrt{4\eta^2+\delta^2}$) as shown in Fig.~\ref{fig2}(a), which is similar to the dynamical behavior of the second-order correlation function.  


In Fig.~3(b), we show the deviations of photon number distribution with respect to the Poisson distribution, which is defined as $\Delta\mathcal{P}_n/\mathcal P_n=(\text{P}_n-\mathcal P_n)/\mathcal P_n$. Here, $\text P_n$ and $\mathcal P_n$ represent the photon distribution of the system and the Poisson distribution with the same mean photon number, respectively. The deviations in photon number distributions show the evidence of the nonclassical field. It is clear to see that the probability of detecting more than two photons at the same time are significantly reduced, while two-photon emission is enhanced compared to a Poissonian field of the same mean photon number. 

 
\begin{figure}[htb]
	\centering
	\includegraphics[width=0.5\textwidth]{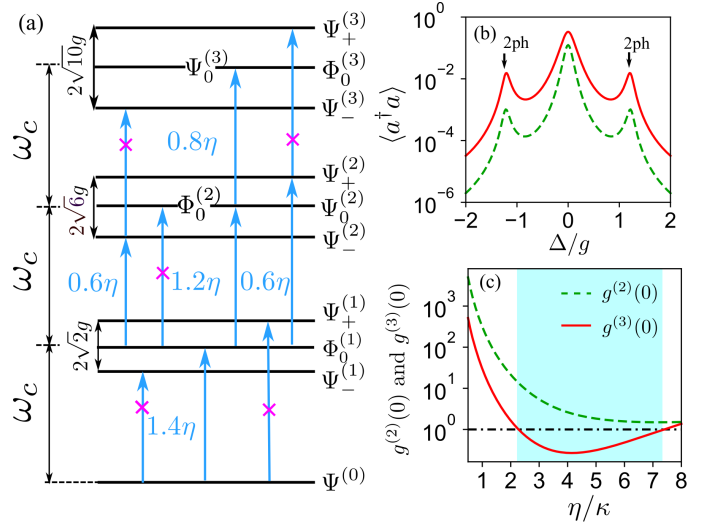}
	\caption{(a) Diagram of eigenstates and the main transition pathways for $\phi_z=\pi$.
		(b) The mean photon number versus the normalizing detuning $\Delta/g$ with $\eta/\kappa=1$ (green line) and $\eta/\kappa=2$ (red line). (c) Steady state field correlation functions $g^{(2)}(0)$ (green line) and $g^{(3)}(0)$ (red line) versus the pump field Rabi frequency $\eta$ at two-photon resonance frequency.	
		}~\label{fig3}
\end{figure}
\section{Out phase radiations.}
Now, we consider the case of $\phi_z=\pi$, where two atoms radiate in different phases. Using the same collective states as basis, the Hamiltonian in $n$-photon space can be expressed as 
\begin{equation}
H_I=
\begin{bmatrix}
0          & 0 & \sqrt{2n}g   &              0\\
0          & 0 & 0             &             0\\
\sqrt{2n}g & 0 & 0             &\sqrt{2(n-1)}g\\
0          & 0 &\sqrt{2(n-1)}g &             0
\end{bmatrix},
\end{equation}
which yields a set of eigenvalues and the corresponding eigenstates shown in Fig.~4(a). In one-photon space ($n=1$), neglecting the state $|ee,0\rangle$, we can obtain the eigenvalues $E^{(1)}_\pm=\pm\sqrt{2}g$ and $E^{(1)}_{0\pm}=0$. The corresponding eigenstates are given by $\Psi_{\pm}^{(1)}=(-|gg,1\rangle\mp|-,0\rangle)/\sqrt{2}$ and $\Phi_{0}^{(1)}=|+,0\rangle$, respectively. Contrary to the case of $\phi_z=0$, the $|gg,0\rangle\leftrightarrow\Phi_0^{(1)}$ transition is allowed so that there exist a resonant peak at the central frequency of the pump field ($\Delta=0$) in the cavity excitation spectrum shown in Fig.~4(b). On the other hand, another two one-photon excitation states are prohibited due to the destructive interference. 
In two-photon space ($n=2$), the eigenvalues are $E^{(2)}_\pm=\pm\sqrt{6}g$ and $E^{(2)}_{0\pm}=0$, and the corresponding eigenstates are given by $\Psi_{\pm}^{(2)}=\sqrt3/3|gg,2\rangle\pm\sqrt2/2|-,1\rangle+\sqrt6/6|ee,0\rangle$, $\Phi_{0}^{(2)}=(-|gg,2\rangle+\sqrt{2}|ee,0\rangle)/\sqrt{3}$ and $\Psi_{0}^{(2)}=|+,1\rangle$, respectively. Increasing the pump field Rabi frequency, two photon excitations are dominant and strong enough to be observed as shown in Fig.~4(b). In three-photon space ($n=3$), we can find the eigenvalues as $E^{(3)}_\pm=\pm\sqrt{10}g$ and $E^{(3)}_{0\pm}=0$ with eigenstates $\Psi_{\pm}^{(3)}=\sqrt{30}/10|gg,3\rangle\pm\sqrt2/2|-,2\rangle+\sqrt5/5|ee,1\rangle$, $\Phi^{(3)}_{0}=(-\sqrt{2}|gg,3\rangle+\sqrt{3}|ee,1\rangle)\sqrt{5}$ and $\Psi^{(3)}_{0}=|+,2\rangle$, respectively. Since the one-photon excitations are forbidden, the three-photon blockade ($g^{(2)}(0)>1$ and $g^{(3)}(0)<1$) can be observed over a wide range of pump field intensity [see Fig.~4(c)]. 




\begin{figure}[htbp]
	\centering
	\includegraphics[width=0.5\textwidth]{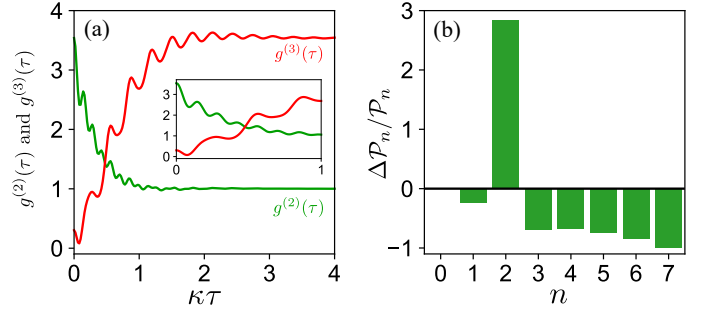}
	\caption{(a) Time dependent field correlation functions $g^{(2)}(\tau)$ (green line) and $g^{(3)}(\tau)$ (red line) versus the normalized delay time $\kappa\tau$ at $\Delta=-\sqrt{6}g/2$ with $\eta/\kappa=3.5$ and $\phi_z=\pi$. (b) The deviations of photon number distribution with respect to the Poisson distribution.
	}~\label{fig4}
\end{figure}
In Fig.~5(a), we plot the time dependent field correlation functions versus the normalized delay time $\kappa\tau$ with $\Delta=-\sqrt{6}g/2$. Compared with the case of $\phi_z=0$, the dynamical properties of the correlation functions are different. Since the one-photon excitations are forbidden and the frequency of the pump field is resonant to the two-photon excitation, the steady state of the system is close to the eigenstate $\Psi_-^{(2)}$. After detecting one photon, the system is projected into the state $|\phi\rangle=\sqrt6/3|gg,1\rangle-\sqrt 2/2|-,0\rangle$. It is noted that the $|gg,1\rangle\overset{\eta}\leftrightarrow|+,1\rangle$ transition is far off-resonant and $|-,0\rangle$ state doesn't coupled to the pump field so that the second-order correlation function oscillates with frequency $2\sqrt{2}g$ due to the $|gg,1\rangle\overset{g_-}\leftrightarrow|-,0\rangle$ transition [see Fig.~5(a), green curve].
For the third-order correlation function $g^{(3)}(\tau)$, the state $\Psi_-^{(2)}$ is still dominant since higher excitation states are off-resonant. After detecting two photons, the system is projected into the state $|gg,0\rangle$ so that the third-order correlation function $g^{(3)}(\tau)$ oscillates with frequency of $\sqrt{4(\sqrt2\eta)^2+\Delta^2}\approx21\kappa$ (corresponding to the time period $T=0.3/\kappa$) due to the $|gg,0\rangle\overset{\eta}\leftrightarrow\Phi_0^{(1)}$ transition [see Fig.~5(a), red curve]. 


Likewise, the deviations of photon number distribution with respect to the Poisson distribution are demonstrated in Fig.~\ref{fig4}(b). It is clear to see that the probability of detecting more than two photons at the same time are significantly reduced as the same as the case of in-phase radiations. However, two-photon emission is significantly enhanced compared to a Poissonian field of the same mean photon number, which can't be observed in the case of in-phase radiations.

\section{Conclusion.}
In summary, we have studied the dynamical properties of the three-photon blockade in a two atoms cavity QED system. We show that, when two atoms radiate in-phase, the three-photon blockade can only be observed in a narrow regime of the pump field Rabi frequency. However, in the case of out-of-phase radiations, the three-photon blockade can be observed in a wide Rabi frequency regime due to prohibiting of the one-photon excitations. Exploring the evolution of the field correlation functions, we show that it is possible to realize the anti-bunched photon pairs in our system when two atoms radiate out-of-phase. 


\begin{acknowledgments}
We acknowledge the National Key Basic Research Special Foundation (Grant No. 2016YFA0302800); the Shanghai Science and Technology Committee (Grant No. 18JC1410900); the National Nature Science Foundation (Grant No. 11774262).
\end{acknowledgments}

\bibliography{reference}

\end{document}